\documentstyle[12pt]{crnart12}
\def\beq{\begin{equation}}
\def\eeq{\end{equation}}
\def\bea{\begin{eqnarray}}
\def\eea{\end{eqnarray}}
\def\bq{\begin{quote}}
\def\eq{\end{quote}}
\catcode`\@=11
\@addtoreset{equation}{section}

\begin{document}
\renewcommand{\thefootnote}{\dag}
\pagestyle{empty}
\vspace*{-.5in}
\begin{flushright}
{CERN-TH.6824/93}\\
{ACT-3/93}\\
{CTP-TAMU-12/93}\\
{SHEP 92/93-14}
\end{flushright}
%\vspace*{5mm}
\begin{center}
{\bf PERTURBATIVE QCD DATA ARE CONSISTENT WITH LIGHT GLUINOS}\\
\vspace*{1cm} {\bf John Ellis} \\
\vspace*{0.3cm}
{\it Theoretical Physics Division, CERN} \\
{\it CH - 1211 Geneva 23} \\
\vspace*{0.3cm}
{\bf D.V. Nanopoulos}\\
{\it Center for Theoretical
Physics, Department of Physics, Texas A\&M University}\\
{\it College Station, TX 77843-4242, USA}
 and
 {\it Astroparticle Physics
Group, Houston Advanced Research Center (HARC)}\\
{\it The Woodlands, TX 77382, USA}\\
\vspace*{0.3cm}
and \\
\vspace*{0.3cm}
{\bf Douglas A. Ross}\\
{\it Physics Department, University of Southampton} \\
{\it Highfield, Southampton SO9 5NH, England} \\
\vspace*{1cm}
{\bf ABSTRACT} \\ \end{center}
%\vspace*{5mm}
\noindent
We discuss systematically the consistency of light gluinos with
data on perturbative QCD from deep inelastic scattering,
quarkonia, jets at LEP, and the total hadronic cross-section in
${e^+ e^-}$ annihilation on the $Z$ peak and elsewhere. We
demonstrate that, in addition to the well-known increase in the
value of $\alpha_s(m_Z)$ inferred from lower-energy data due
to the slower running of $\alpha_s$ in the presence of light
gluinos, the value of $\alpha_s(m_Z)$ extracted from the LEP
data must also be $increased$, as a result of including the
effects of virtual light gluinos. The effect of these increases
in other estimates of $\alpha_s(m_Z)$ is to make them more
consistent with the value extracted from the total $e^+ e^-$
cross-section, which would otherwise appear distinctly higher.
We discuss the possibility of looking for light gluinos in
four-jet events at LEP, and their possible implications for
scaling violations at HERA.

\vspace*{0.3cm}

\begin{flushleft}
CERN-TH.6824/93 \\
{ACT-3/93}\\
{CTP-TAMU-12/93}\\
{SHEP 92/93-14}\\
March 1993
\end{flushleft}
\vfill\eject
%\pagestyle{empty}
%\clearpage\mbox{}\clearpage

\setcounter{page}{1}
\pagestyle{plain}
 {\newcommand{\la}{\mbox{\raisebox{-.6ex}{$\stackrel{<}{\sim}$}}}
{\newcommand{\ga}{\mbox{\raisebox{-.6ex}{$\stackrel{>}{\sim}$}}}
\voffset -1in
\textheight=20cm
\vskip2.0cm

\section{Introduction}

It is normally assumed in discussions of supersymmetric
phenomenology that all the sparticles have masses so large
that they are beyond the physics reach of accelerators now
running. However, there is one persistent possible exception to this
general assumption, namely that the gluinos are very light,
weighing at most a few $GeV$ \cite{FF}.
This possibility has been severely
constrained by a number of different experimental searches,
notably at the CERN $p \bar p$ collider \cite{UA1},
in bottomonium decays,
and in fixed-target experiments
looking for metastable particles or missing
energy \cite{HELIOS},\cite{NA3}.
Nevertheless, there are two possible windows for light
gluino masses and lifetimes that do not seem to be ruled out
by these analyses. One is for gluinos weighing between 3 and
4 $GeV$ with lifetimes around $10^{-13} s$, and the other is
for gluinos weighing 3 $GeV$ or more and having lifetimes
between about $10^{-8}$ and $10^{-10} s$.

Two of us (J.E. and D.V.N.) argued some time ago together
with I. Antoniadis \cite{AEN} that these windows could be explored
indirectly via the indirect effects of light gluinos in loop
diagrams. Specifically, we pointed out that the consistency of
low-energy determinations of $\alpha_s$, for example in deep
inelastic scattering, quarkonium or $\tau$ decays, with high-energy
determinations of $\alpha_s(m_Z)$ constrains in principle the
contributions of light coloured particles to the renormalization
group running of $\alpha_s$, and in particular the possible
contributions of light gluinos. We showed that the LEP jet data
then available  disfavoured light gluinos,
but we said that ``This conclusion should be regarded as
preliminary, pending further reduction of the
experimental and theoretical uncertainties".
Subsequently,
the LEP jet data and their theoretical understanding have advanced
somewhat, and the preferred value of $\alpha_s$ extracted from
LEP jet data has increased significantly. This has led to renewed
questioning whether light gluinos are consistent with the LEP and
other perturbative QCD data \cite{Clavelli},\cite{JK}.

In fact, as we show later in this paper, one cannot take blindly
the values of $\alpha_s(m_Z)$ extracted from LEP jet data and
compare them directly with lower-energy data extrapolated with
the slower
running of $\alpha_s$ caused by the presence of light gluinos.
This is because the LEP values of $\alpha_s(m_Z)$ have been
extracted from analyses using loop corrections \cite{ERT}
in which the
possible effects of light gluinos have not been included. Any
analysis claiming consistency of the perturbative QCD data
must include light gluinos consistently in all loop diagrams,
not just in running $\alpha_s$ up from the typical energies
of deep inelastic scattering, quarkonium or $\tau$ decays to
the energy of LEP.

The main purpose of this paper is precisely to furnish such a
consistent analysis, including gluino loop effects in the
analysis of LEP jet data and the total hadronic cross-section
on the $Z$ peak \footnote{This was not done in \cite{Clavelli},
\cite{JK}.}.
We find that including the virtual effects of
light gluinos $increases$ the jet values of $\alpha_s(m_Z)$ by
about $10$ percent, and also increases the value inferred from the
total cross-section at LEP, although by less than $2$ percent. The net
effect of these increases, combined with the change in the running
of $\alpha_s$ up from lower energies, is that the overall
consistency of perturbative determinations of $\alpha_s(m_Z)$ is
$improved$ if light gluinos are present, since the other
determinations of $\alpha_s(m_Z)$ are now in better agreement with
that extracted from the total cross-section. However, this
improvement is not very significant, and we do not claim
that the data are inconsistent with the absence of light gluinos.
Nevertheless, we feel that our analysis reinforces the
desirability of further direct searches for light gluinos. One
way to do this is by examining the rate and characteristics of
four-jet events at LEP. Another is by careful analysis of scaling
violations at HERA, where the effective value of $\alpha_s$
would be about $10$ percent
 higher in the presence of light gluinos
than without them, and real gluinos might be observable at low
values of the Bjorken scaling variable $x$.

\section{The Running of $\alpha_s$ from Low Energies}

As a warm-up exercise, we first discuss the running of the
low-energy $\alpha_s$ data up to $m_Z$. The one-loop expression
for the QCD $\beta$ function \footnote{This is defined by
$$ \frac{d\alpha_s(Q^2)}{d\log (Q^2)}= \beta \alpha_s(Q^2)$$ }
 including light gluinos is
\beq   -\frac{\alpha_s}{4\pi} \left(9-\frac{4n_f}{3} \right)
\label{b_1coefficient}
\eeq
and the two-loop expression is
\beq   -\left( \frac{\alpha_s}{4\pi} \right)^2 \left( 54 - \frac{38n_f}{3}
\right)
 \label{b_2coefficient} \eeq
where $n_f$ is the number of flavours.

% threshold at twice the gluino mass
These expressions are to be used for momenta $Q$ larger than
the threshold for gluino pair production, i.e. twice the
gluino mass $m_{gluino}$, which we shall vary over the range
$3$ to $5$ $GeV$. The most significant low-energy data are those
extracted from $\tau$ decays, charmonium spectroscopy, bottomonium
decays, and deep inelastic scattering in fixed-target $\nu$ and $\mu$
experiments \cite{Bethke}.
The first two of these are at $Q$ values below
$m_{gluino}$, whilst the gluino threshold might be within the
kinematic ranges of the latter two. The relevant average
momentum transfers
in the
deep inelastic data are about $5$ $GeV$ for $\nu$ scattering
and $7.1$ $GeV$ for $\mu$ scattering. We list in Table $1$ the values
of $\alpha_s(m_Z)$ extracted from the low-energy data,
both without gluinos, as reviewed
in ref. \cite{Bethke},  and with gluinos of
various different masses. We see that the effect of including
gluinos is to increase the extrapolated value of $\alpha_s(m_Z)$
by as much as $15$ percent. Since the low-energy data in the absence
of gluinos tend to indicate a lower value of $\alpha_s(m_Z)$ than
that inferred from LEP data, this increase is promising, but we
re-analyze the LEP data themselves before drawing any
conclusions.

\section{Virtual Gluino Effects on LEP Jet Cross-sections}

In this section we shall mainly be concerned with the three-jet
cross-section, which is given by \cite{EGR}
\beq \frac{1}{\sigma_0}d\sigma_0^{(3)}(y_{13},y_{23}) \ = \
\ \frac{2\alpha_s}{3\pi} \frac{(1-y_{13})^2+(1-y_{23})^2}{y_{13}y_{23}}
\label{firstDouglas} \eeq
at the tree level, where $\sigma_0$ is the leading-order contribution
to the total cross-section, and $y_{ij}$ = ${(p_i - p_j)}^2/{m_Z}^2$.
The three-jet cross-section is given at the one-loop level \cite{ERT}
by an
expression of the form
\beq
\frac{1}{\sigma_0}d\sigma_1^{(3)}(y_{13},y_{23}) \ = \
\frac{1}{\sigma_0}d\sigma_0^{(3)}(y_{13},y_{23})
\frac{\alpha_s}{2\pi} \left[ \frac{17}{6} \pi^2 +
\rho(y_{13},y_{23}) \right]
\label{secondDouglas}\eeq
where we have pulled out explicitly a correction term proportional
to $\pi^2$, and the form of the residual correction
$\rho(y_{13},y_{23})$ will be discussed shortly. We assume that
the infrared divergence in the three-jet cross-section has been
regularized by combining it with the infrared part of the
four-jet cross-section $d\sigma^{(4)}/\sigma_0$, which is also of
order ${\alpha_s}^2$.

We now consider an arbitrary event shape variable $X$, which is
given for three-jet final states by $X^{(3)}(y_{12},y_{23})$, and
for four-jet final states by $X^{(4)}(....)$. The corresponding
differential cross-section is then given by
\bea
\frac{1}{\sigma_0} \frac{d\sigma}{dX} \ & \ = &
\ \int dy_{13}dy_{23} \theta(1-y_{13}-y_{23}) \delta(X-X^{(3)}
(y_{13},y_{23}))
\frac{d\sigma_0^{(3)}(y_{13},y_{23})}{\sigma_0} \nonumber \\
 & & \times \left\{ 1+\frac{\alpha_s}{2\pi}\left[
 \frac{17}{6} \pi^2 + \rho (y_{13},y_{23}) \right] \right\}
 \nonumber \\ & &
+ \int d^{(4)}LIPS \frac{d\sigma_0^{(4)}}{\sigma_0}\delta(X-X^{4}(...))
\label{thirdDouglas}
\eea
and  the generic correction factor, $\eta(X)$, defined in ref.\cite{ENR}
is given by
\bea
 6 \eta (X) \ \int dy_{13}dy_{23} \theta(1-y_{13}-y_{23})
\delta(X-X^{(3)}(y_{13},y_{23}))
\frac{d\sigma_0^{(3)}(y_{13},y_{23})}{\sigma_0} & = & \nonumber \\
\ \int dy_{13}dy_{23} \theta(1-y_{13}-y_{23})
\delta(X-X^{(3)}(y_{13},y_{23}))
\frac{d\sigma_0^{(3)}(y_{13},y_{23})}{\sigma_0}\rho (y_{13},y_{23}) & &
\nonumber \\
  + \frac{2\pi}{\alpha_s}
 \int d^{(4)}LIPS \frac{d\sigma_0^{(4)}}{\sigma_0}
 \delta(X-X^{4}(...)) & &
\label{fourthDouglas}
\eea
To an extremely good approximation, as good as the accuracy of the
% enlarge on the accuracy of the approximation, THROUGHOUT 3-jet region
Monte Carlo routine used to perform the phase space integral for the
higher order corrections,
 the part of the higher-order correction (\ref{secondDouglas})
 that depends on the
light fermions is given throughout the three-jet region by
\beq
\rho^{Fermi}(y_{13},y_{23}) \ =
 \ T_R \left( \frac{2}{3} \log (y_{13}y_{23})-\frac{10}{9} \right)
\label{fifthDouglas}\eeq
where quarks contribute $n_f/2$ to $T_R$,
and we can neglect $e^+e^- \rightarrow q{\bar q}q^\prime{\bar q}^\prime$,
so that $d\sigma^{(4)}$ = $0$ throughout the three-jet kinematic
range. For the case of the Fox-Wolfram variable, $C$,
the accuracy of this
approximation can be seen by a direct comparison with the fermionic part
of the higher-order correction displayed explicitly in \cite{ERT}.
The key point for this analysis is that whereas
the gluon corrections to the three jet cross section are positive the
fermion-dependent contribution (\ref{fifthDouglas}) is
{\it negative}. This  means
that, when $T_R$ is increased by the appearance of some new species of
fermion such as a light gluino,
the one-loop correction to the three-jet cross-section is
{\it decreased} by virtual fermion effects. This phenomenon is
exhibited in the figure for the thrust variable $T$ and the
Fox-Wolfram variable $C$. It is evident, therefore, that the
value of ${\alpha_s(m_Z)}$ extracted from a fit to such a
distribution will be {\it increased}.

Before analysing this effect numerically, we first make some comments
about the mathematical formulae to be used in the fits. We have
argued previously \cite{ENR}
that the $\pi^2$ terms in the one-loop
correction (\ref{secondDouglas}) are likely to exponentiate
over most of the
three-jet kinematic range when higher orders are calculated. We
are well aware that this exponentiation has not been proved, but
we assume it here as a working hypothesis. In this case, the
one-loop expression becomes
\bea
\frac{1}{\sigma_0} \frac{d\sigma}{dX} \ & \ = &
\ \int dy_{13}dy_{23} \theta(1-y_{13}-y_{23})
\delta(X-X^{(3)}(y_{13},y_{23}))
\frac{d\sigma_0^{(3)}(y_{13},y_{23})}{\sigma_0} \nonumber \\
 & & \times \left\{ \exp \left(
\frac{17\alpha_s\pi}{12}  \right) \left[ 1
 + \frac{\alpha_s}{2\pi}\rho (y_{13},y_{23}) \right] \right\}
 \nonumber \\ & &
+ \int d^{(4)}LIPS \frac{d\sigma_0^{(4)}}{\sigma_0}\delta(X-X^{4}(...))
\label{sixthDouglas}
\eea
The recipe for including gluinos is simply to add to
$\rho(y_{13},y_{23})$ the expression (\ref{fifthDouglas}) for
$\rho^{Fermi}$
with an extra contribution to $T_R$ of $3/2$,
so that the differential cross-section
becomes $d\sigma^{w.g.}$, given by
\bea
\frac{1}{\sigma_0} \frac{d\sigma^{w.g.}}{dX} \ & \ = &
\ \int dy_{13}dy_{23} \theta(1-y_{13}-y_{23})
\delta(X-X^{(3)}(y_{13},y_{23}))
\frac{d\sigma_0^{(3)}(y_{13},y_{23})}{\sigma_0} \nonumber \\
 & & \hspace{-1in}\times \left\{ \exp \left(
\frac{17\alpha_s\pi}{12}  \right) \left[ 1
 + \frac{\alpha_s}{2\pi} \left( \rho (y_{13},y_{23})+\log (y_{13}y_{23})
  -\frac{5}{3} \right) \right] \right\} \nonumber \\ & &
+ \int d^{(4)}LIPS \frac{d\sigma_0^{(4)}}{\sigma_0}\delta(X-X^{4}(...))
\label{seventhDouglas}
\eea
However, there is an ambiguity in this procedure: should the
exponential factor also appear in front of the non-$\pi^2$
correction term $\rho(y_{13},y_{23})$? No-one knows, and we
consider this as a theoretical uncertainty in the
extraction of $\alpha_s(m_Z)$. Thus, we consider in our
subsequent numerical results two possible expressions for the
gluino corrections to the differential cross-section for a
generic three-jet variable: one without exponentiation of the
gluino correction:
\bea
\frac{1}{\sigma_0} \frac{d\sigma^{w.g.}}{dX}  \  & \ =  \ & \
\frac{1}{\sigma_0} \frac{d\sigma}{dX} +
  \frac{\alpha_s}{2\pi}
\ \int dy_{13}dy_{23} \theta(1-y_{13}-y_{23})
\delta(X-X^{(3)}(y_{13},y_{23}))
\nonumber \\ & &
 \times \frac{d\sigma_0^{(3)}(y_{13},y_{23})}{\sigma_0}
\left( \log(y_{13}y_{23})-\frac{5}{3} \right)
\label{eighthDouglas}
\eea
and one with exponentiation of the gluino correction:
\bea
\frac{1}{\sigma_0} \frac{d\sigma^{w.g.}}{dX}  \  & \ =  \ & \
\frac{1}{\sigma_0} \frac{d\sigma}{dX} +
  \frac{\alpha_s}{2\pi}
\ \int dy_{13}dy_{23} \theta(1-y_{13}-y_{23})
\delta(X-X^{(3)}(y_{13},y_{23}))
\nonumber \\ & &
 \hspace{-.5in} \times  \exp \left(
\frac{17\alpha_s\pi}{12}  \right)
\frac{d\sigma_0^{(3)}(y_{13},y_{23})}{\sigma_0}
\left( \log(y_{13}y_{23})-\frac{5}{3} \right)
\label{ninthDouglas}
\eea
where the first term on the R.H.S. in each case is the differential
cross section in the absence of light gluinos.

%Mention table later, point out that thick line measures error
In the histograms shown in the Figure
which display the effects of including
light gluinos, the thickness of the solid line is a measure of the
uncertainties in the effects,
given by the difference between the two above  expressions.

% explain M_D a bit
We have studied the following three-jet variables: thrust $T$,
oblateness $O$, energy-energy correlations $EEC$, the asymmetry
in energy-energy correlations $AEEC$, the Fox-Wolfram variable
$C$, the heavy jet invariant mass $M_H$, and the quantity $M_D$,
which measures the difference between the light and heavy
jet invariant masses. The
experimental values of these quantities have been taken from the
review \cite{Bethke}. The corresponding values of $\alpha_s(m_Z)$
extracted assuming the presence and absence of gluinos are shown
in Table 2. In each case, the ranges quoted for the estimates
with gluinos accounts for the above-mentioned ambiguity, and
the gluino mass is assumed to be much lighter than $M_Z$.
 We see that the inclusion
of gluinos increases $\alpha_s(M_Z)$ by about $10$ percent. The
bottom row of Table 2 gives the weighted average values of
$\alpha_s(M_Z)$. For the reason given in ref. \cite{ENR}, namely
that the non-$\pi^2$ correction $\eta$ (\ref{fourthDouglas})
is particularly large
for these variables, we prefer to exclude oblateness and the $AEEC$
from the weighted average, but we {\it do}
include the effect of exponentiation
of the $\pi^2$ part of the correction factor discussed in
\cite{ENR}. This has the
effect of reducing the average value of $\alpha_s(M_Z)$,
as shown in the last line of
Table 2. We find that in the presence of light gluinos
\beq
\alpha_s(M_Z) \ = \  0.124 \pm .007
\label{bottomline} \eeq
%say something positive about the error estimate
for the value of $\alpha_s(M_Z)$ extracted from
weighted averages of three-jet data on the $Z$ peak. The error is
obtained from the errors quoted in \cite{Bethke} added in quadrature with
the theoretical  error on the effect of the gluinos.
This is to be compared with the value of $\alpha_s(m_Z)=0.113 \pm$.006
in the absence of gluinos but with $\pi^2$ exponentiation. The
penultimate line of Table 2 gives the corresponding values of
$\alpha_s(m_Z)$ with and withour gluinos, but without $\pi^2$
exponentiation.

\section{Total $e^+ e^-$ Cross-section Data}

These can be used to extract a value of $\alpha_s(M_Z)$ using the
next-to-leading order formula \cite{chet}
\beq
\sigma^{tot}(e^+e^-)=\sigma_0^{tot}(e^+e^-)
\left[ 1 + \frac{\alpha_s}{\pi}+\left( \frac{\alpha_s}{\pi}\right)^2
(1.98-0.230T_R) \right]
\eeq
The available data are well known to give a value of $\alpha_s(M_Z)$
that is consistently higher than other determinations, though only
about one standard deviation higher than the latest three-jet
measurements on the Z peak, in the absence of light gluinos. To
extract the value of $\alpha_s(M_Z)$ in the presence of light
gluinos, we can to this order simply add three more effective flavours
and equate
\beq
\frac{\alpha_s(M_Z)}{\pi}+1.41\left(
\frac{\alpha_s(M_Z)}{\pi}\right)^2 \ =
 \ \frac{\alpha_s^\prime(M_Z)}{\pi}+1.06 \left(
\frac{\alpha_s^\prime(M_Z)}{\pi}\right)^2
\label{backofenvelope}
\eeq
where $\alpha_s^\prime(M_Z)$ is the value which would be extracted form
the total cross-section if gluinos were included, and $\alpha_s(M_Z)$
is the value of $0.130\pm.012$ quoted in \cite{Bethke}. This yields
\beq
\alpha_s^\prime(M_Z) \ = \ 0.132 \pm.012
\label{whatever}
\eeq
% enlarge on smaller effect.
Thus we find yet another
increase in the extracted value of $\alpha_s(M_Z)$, but by less
than $2$ percent this time. For the total cross-section the
higher-order corrections are smaller (both for the gluon and
fermion corrections)
because of a partial cancellation by the two-loop correction to the
two-jet cross-section. In fact it can be seen from (\ref{fifthDouglas})
that the effect of adding light gluinos actually diverges
logarithmically as one goes to the two-jet region ($y_{13}\rightarrow 0$
or $y_{23}\rightarrow 0$), and this divergence is cancelled by an
infrared-divergent correction to the two-jet cross-section.\footnote{The
divergence is actually regulated by the gluino mass,
but for light gluinos
 we can neglect this for energies of order $M_Z$.}. The net effect
is a reduction in the higher-order correction for the total cross-section
compared with the jet event shape differential cross-sections.

% a bit more positive about the improvement of the consistency
The overall consistency of the available data on $\alpha_s(M_Z)$ is
therefore
improved by the inclusion of light gluinos, in that the spread of
central values is reduced from $0.112$-$0.130$ to $0.124$-$0.136$,
but this effect is
not conclusive. The most significant effect of virtual gluinos is
not to reconcile the low-energy and LEP jet data, whose central
values are not brought
closer together ($0.124$-$0.136$ compared with $0.112$-$0.121$
previously). We suspect that this is because the three-jet
data effectively measure the $\alpha_s$ at a momentum scale
significantly below $M_Z$. On the other hand, including light
gluinos does bring other determinations of $\alpha_s(M_Z)$ into
better agreement with the value extracted from the total hadronic
cross-section at the $Z$ peak, in that a maximum possible
discrepancy of $0.112$ vs $0.130$ is reduced to $0.124$ vs $0.132$.
It has been a long-standing puzzle
that total cross-section measurements consistently give larger
values of $\alpha_s$, although the difference has never been more
than one standard deviation or so. Thus, there has never really
been a problem. Nevertheless the improved agreement is interesting
and may become significant as the statistical errors on
the measurement of $\alpha_s(M_Z)$ are further reduced. Even though
 there must be less exotic solutions to the potential problem of a
mismatch in the values of $\alpha_s(M_Z)$ extracted from different
data, the fact that data on the running of $\alpha_s$ from low energies
to $M_Z$ cannot rule out light gluinos,
together with the improved consistency of the LEP data
when gluinos are included, re-awakens our interest
in possible direct searches.

\section{Search for Light Gluinos in Four-Jet Events at LEP}

%ERT formula much too long to reproduce here, - just refer
It was suggested several years ago \cite{CER} to look for light gluinos
in four-jet events in $e^+ e^-$ annihilation, produced by gluon
splitting after bremsstrahlung from an initial $q$ or ${\bar q}$.
Neglecting the gluino mass, which should be a good approximation
for events passing the selection criteria of the LEP
experiments, the differential cross-section is proportional to
that for $q$${\bar q}$$q'$${\bar q'}$ final states in $e^+e^-$
annihilation,
where $q$ and $q'$ are both light flavours.
The differential cross-section for this process can be found in
\cite{ERT}.
Note that, as mentioned in ref. \cite{CER}, there is a
factor of $3$ relative to the cross-section calculated
for light quarks
in ref. \cite{ERT}: this is composed of a factor $6$ for colour,
and $1/2$ for the Majorana nature of the gluinos.

Two of the
LEP collaborations have recently published \cite{DELPHI,ALEPH}
detailed analyses of
four-jet final states in which they make a kinematic separation
of the contributions due to double gluon bremsstrahlung, gluon
splitting to gluons, and gluon splitting to fermions. These
studies were motivated by the wish to measure independently the
colour charges of gluons and quarks, and the results were
consistent with the Standard Model. In the presence of light
gluinos, the value of $T_R=n_f/2$ for quarks measured in these
experiments would be enhanced to
% Surely we must mean 3/2 not 3/10 !!
\beq
n_f/2 + 3/2
\eeq
The data published so far are also consistent with this enhanced
value.

Discriminating between the Standard Model and the
presence of light gluinos will require further work on at least
two fronts: the QCD radiative corrections to the four-jet
cross-section should be evaluated, and the distinctive properties
of gluino jets should be investigated. Gluinos are expected to
decay into invisible, weakly-interacting neutral particles that
carry away missing energy. Any experiment looking at four-jet
events needs to consider carefully whether such an energy loss
could bias the jet energy determination, possibly
shifting gluino events
to lower apparent energies where there is more background, or
even pushing gluino jets below the experimental cuts. On the
other hand, gluinos in the longer-lived part of the light gluino
window, namely with a lifetime between $10^{-8}$ and $10^{-10}$ $s$,
might yield events with detectable separated vertices in the
four-jet region.

We therefore think that the light gluino window could be closed
(or opened) by determined searches at LEP.

\section{Light Gluino Effects at HERA}

If gluinos are indeed light, they could have both indirect and
direct effects at HERA. As we have discussed above, the preferred
value of $\alpha_s(M_Z)$ extracted from LEP and lower-energy
data is increased by about $10$ percent, to around $0.13$,
if light gluinos are present.\footnote{Although it lies beyond
the scope of this paper, we note that such a large value of
$\alpha_s(m_Z)$would endanger the high degree of consistency of
LEP data \cite{EKN} with minimal supersymmetric GUTs.}
This difference should show up
clearly as enhanced scaling violations at large momentum
transfers at HERA, due to the reduction in the value of $\beta$.
 In addition to this indirect effect, there
could also be an observable direct effect via the production
of real light gluino pairs at small values of the Bjorken
scaling variable $x$. Their production would be a higher-order
QCD effect whose evaluation, as well
as that of the conventional low-$x$
background, goes beyond the scope of this paper. However, we
recall that, as discussed in case of four-jet events at LEP,
gluino pair production events could have distinctive missing
energies and possibly separated vertices.

We therefore think that the light gluino window could also be
closed (or opened) by thorough searches at HERA.

{\bf Acknowledgements}

One of us (JE) thanks the University of Southampton for its kind
hospitality while this work was initiated. The work of DVN was
supported in part by DOE grant DE-FG05-91-ER-40633 and by a grant
from Conoco Inc.
}
}

\newpage
\begin{center} {\bf Table 1} \\
  Extraction of $\alpha_s(M_Z)$ from low energy data \end{center}
\hspace*{-.7in}
\begin{tabular}{||c|l|l|l|l ||} \hline \hline
 &  \multicolumn{4}{c||}{}\\
Process &   \multicolumn{4}{c||}{$\alpha_s(M_Z)$} \\ \cline{2-5}
  & Without gluinos & $m_{gluino}=3 \ GeV$ & $m_{gluino}=4 \ GeV$
  & $m_{gluino}=5 \ GeV$ \\
\hline \hline
  & & & & \\
 $\frac{\Gamma(\Upsilon \rightarrow ggg)}
{\Gamma(\Upsilon \rightarrow \mu^+\mu^-)}_{(LEP)}$ &
 $0.121\pm .005$ & $0.140\pm .006$ & $0.137\pm
 0.006$ & $0.136\pm .006$ \\
  & & & & \\
$\frac{\Gamma(\Upsilon \rightarrow ggg)}
{\Gamma(\Upsilon \rightarrow \mu^+\mu^-)}_{(World)}$ &
 $0.118\pm .005$ & $0.136\pm .006$ & $0.134
 \pm 0.006$ & $0.132\pm .006$ \\
  & & & & \\ \hline
  & & & &  \\
 $J/\Psi, \  \Upsilon$ decays & $0.113\pm .005$ & $ 0.129\pm .006$ &
 $0.127\pm .006$ & $0.126\pm .006$ \\
 & & & & \\ \hline
 & & & &  \\
 D.I.S. & $0.112\pm .004$ & $0.129 \pm .005$ & $0.127
 \pm .005 $ & $ 0.125\pm
.005 $ \\
 & & & & \\
 \hline \hline \end{tabular}

\newpage
\begin{center} {\bf Table 2} \\
 Extraction of $\alpha_s(M_Z)$ from jets data at LEP \bigskip \\
\begin{tabular}{||c|l|l||} \hline \hline
 &  \multicolumn{2}{c||}{}\\
Event Shape Variable &   \multicolumn{2}{c||}{$\alpha_s(M_Z)$}
\\ \cline{2-3}
  & Without gluinos & With gluinos \\
\hline \hline
      & 0.120 ($\aleph$)  & 0.129-0.134 \\
 C    & 0.124 (Delphi)    & 0.134-0.139 \\
      & 0.127 (Opal)      & 0.137-0.142 \\ \hline
      & 0.126 ($\aleph$)  & 0.134-0.139 \\
 T    & 0.123 (Delphi)    & 0.131-0.135 \\
      & 0.118 (L3)        & 0.126-0.130 \\
      & 0.127 (Opal)      & 0.135-0.140 \\ \hline
      & 0.126 ($\aleph$)  & 0.143-0.152 \\
 EEC  & 0.129 (Delphi)    & 0.147-0.156 \\
      & 0.128 (Opal)      & 0.152-0.162 \\ \hline
      & 0.132 ($\aleph$)  & 0.141-0.145 \\
$M_H$ & 0.125 (Delphi)    & 0.133-0.137 \\
      & 0.128 (Opal)      & 0.136-0.141 \\ \hline
      & 0.141 ($\aleph$)  & 0.150-0.155 \\
$M_D$ & 0.122 (Delphi)    & 0.130-0.134 \\
      & 0.118 (Opal)      & 0.126-0.130 \\ \hline
      & 0.113 ($\aleph$)  & 0.122-0.127 \\
  O   & 0.122 (Delphi)    & 0.132-0.137 \\
      & 0.121 (Opal)      & 0.130-0.136 \\ \hline
      & 0.110 ($\aleph$)   & 0.114-0.115 \\
 AEEC & 0.114 (Delphi)    & 0.118-0.120 \\
      & 0.116 (Opal)      & 0.120-0.122 \\ \hline\hline
      & & \\
Weighted Average & 0.120$\pm .006$ & 0.132$\pm .007$ \\
      & & \\ \hline \hline
      & & \\
After $\pi^2$ &&\\
Exponentiation & 0.113 $\pm$ .006 & 0.124 $\pm$ .007 \\
      & & \\ \hline\hline
 \end{tabular}
 \end{center}

\newpage \begin{center} {\bf Figure Caption} \end{center}
Histograms of the $C$ and thrust distributions
for the three-jet region, showing the lowest-order
differential cross-section
(dotted line) and the higher-order differential cross section without
gluinos (thin solid line) and with gluinos (thick solid line).
The thickness
of the solid line is a measure of the theoretical uncertainty in the
gluino effect.
\end{document}